\documentclass[%
 reprint,
unsortedaddress,
showpacs,preprintnumbers,
 amsmath,amssymb,
 aps,
prl,
 longbibliography,
 lengthcheck,%
]{revtex4}

\usepackage{graphicx}
\usepackage{dcolumn}
\usepackage{bm}
\usepackage{hyperref}

\begin{document}

\title{Hugoniot of shocked liquid deuterium up to 300 GPa: Quantum molecular dynamic simulations}

\author{Cong Wang$^{1}$}
\author{Xian-Tu He$^{1,2}$}
\author{Ping Zhang$^{1,2,}$}
\thanks {Corresponding author; zhang\_ping@iapcm.ac.cn}
\affiliation{$^{1}$LCP, Institute of Applied Physics and
Computational Mathematics, P.O. Box 8009, Beijing 100088, People's
Republic of China}

\affiliation{$^{2}$Center for Applied Physics and Technology,
Peking University, Beijing 100871, People's Republic of China}

\begin{abstract}
Quantum molecular dynamic (QMD) simulations are introduced to
study the thermophysical properties of liquid deuterium under
shock compression. The principal Hugoniot is determined from the
equation of states, where contributions from molecular
dissociation and atomic ionization are also added onto the QMD
data. At pressures below 100 GPa, our results show that the local
maximum compression ratio of 4.5 can be achieved at 40 GPa, which
is in good agreement with magnetically driven flyer and
convergent-explosive experiments; At the pressure between 100 and
300 GPa, the compression ratio reaches a maximum of 4.95, which
agrees well with recent high power laser-driven experiments. In
addition, the nonmetal-metal transition and optical properties are
also discussed.
\end{abstract}


\maketitle

Recent studies of materials under extreme conditions, which require
improved understandings of the thermophysical properties in the new
and complex regions, have gained much scientific interest
\cite{PBX:Ernstorfer:2009}. The combination of high temperature and
high density defines \textquotedblleft warm dense
matter\textquotedblright\ (WDM) - a strongly correlated state, which
is characterized by partially dissociated, ionized and degenerated
states, and the modelling of the dynamical, electronic, and optical
properties for such system is rather challenging. Due to their
simplicity, hydrogen and its isotopes (deuterium and tritium) have
been studied intensively \cite{PBX:Nellis:2006}, where the relative
pressure and temperature have reached megabar range and several eV.
Specially, as one of the target materials in the inertial
confinement fusion experiments \cite{PBX:Philippe:2010}, deuterium
has been extensively investigated through experimental measurements
and theoretical models. Gas gun \cite{PBX:Nellis:1983}, converging
explosive \cite{PBX:Boriskov:2005}, magnetically driven flyer
\cite{PBX:Knudson:2004}, and high power laser-driven
\cite{PBX:Collins:1998,PBX:Boehly:2004,PBX:Hicks:2009} experiments
have been applied to probe the physical properties of deuterium
during single or multiple dynamic compression. Theoretically,
approximations have been introduced to simulate warm dense deuterium
(hydrogen), such as, linear mixing model \cite{PBX:Ross:1998},
chemical model FVT \cite{PBX:Juranek:2000}, path integral Monte
Carlo (PIMC)
\cite{PBX:Magro:1996,PBX:Militzer:2000,PBX:Bezkrovniy:2004}, and
quantum molecular dynamics \cite{PBX:Lenosky:2000,PBX:Holst:2008}.

To date, although a number of explanatory and predictive results in
some cases have already been provided by experimental and
theoretical studies, however, many fundamental questions of
deuterium under extreme conditions are still yet to be clarified.
The equation of states (EOS), especially the Hugoniot curve, are
essential in this context. Since five to six-fold the initial
densities have been detected by laser-driven experiments at megabar
pressure regime
\cite{PBX:Boehly:2004,PBX:Hicks:2009,PBX:Collins:1998} and supported
by PIMC simulations \cite{PBX:Magro:1996,PBX:Bezkrovniy:2004},
considerable controversies in the deuterium EOS have been raised.
Meanwhile, converging explosives \cite{PBX:Boriskov:2005} and
magnetically driven flyer \cite{PBX:Knudson:2004} experiments
indicate that the compression ratio ($\eta$) shows a maximum close
to 4.3, which is in good agreement with QMD results
\cite{PBX:Lenosky:2000,PBX:Holst:2008}. Furthermore, in adiabatic
and isentropic compressions, the nonmetal-metal transition of
deuterium (hydrogen), which is accompanied by the change of optical
spectroscopies \cite{PBX:Celliers:2000}, has been a major issue
recently. The links between nonmetal-metal transition and
dissociation (ionization) under dynamic compression are of
particular significance \cite{PBX:Nellis:2006}.

The chemical pictures of deuterium under extreme conditions could be
briefly described as two processes: (i) partial dissociation of
molecules, D$_{2}\rightleftarrows2$D, and (ii) a subsequent
ionization of atoms, D$\rightleftarrows e+$D$^{+}$. QMD simulations,
where electrons are modelled by quantum theory, are convinced to be
a powerful tool to describe the chemical reactions, such as
dissociation and recombination of molecules. Meanwhile, the
dynamical, electrical and optical properties of warm dense matter
have already been proved to be successfully investigated by QMD
simulations \cite{PBX:Kietzmann:2008,PBX:Lorenzen:2009}. However,
the ionization of atoms is not well defined in the framework of
density functional theory (DFT). Considering these facts, thus, in
this paper we applied the corrected QMD simulations to shock
compressed deuterium, and the calculated compression ratio is
substantially increased according to the ionization of atoms in the
warm dense fluid.

We have performed simulations for deuterium by employing the Vienna
Ab-initio Simulation Package (VASP)
\cite{PBX:Kresse:1993,PBX:Kresse:1996}. A fixed volume supercell of
$N$ atoms, which is repeated periodically throughout the space,
forms the elements of the calculation. By involving Born-Oppenheimer
approximation, electrons are fully quantum mechanically treated
through plane-wave, finite-temperature DFT \cite{PBX:Lenosky:2000},
and the electronic states are populated according to the Fermi-Dirac
distribution at temperature $T_{e}$. The exchange correlation
functional is determined by generalized gradient approximation (GGA)
with the parametrization of Perdew-Wang 91 \cite{PBX:Perdew:1991}.
The ion-electron interactions are represented by a projector
augmented wave (PAW) pseudopotential \cite{PBX:Blochl:1994}. The
system is calculated with the isokinetic ensemble (NVT), where the
ionic temperature $T_{i}$ is kept constant every time step by
velocity scaling, and the system is kept in local thermodynamical
equilibrium by setting the electron ($T_{e}$) and ion ($T_{i}$)
temperatures to be equal.

\begin{figure}[!htbp]
\centering
\includegraphics[width=6.0cm]{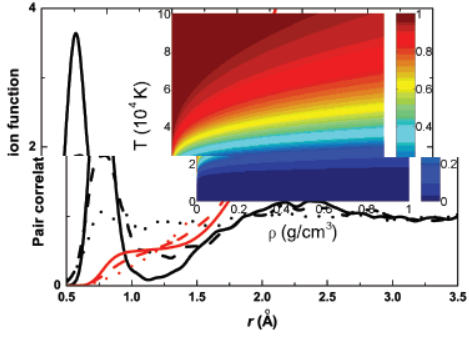}
\caption{(Color online) Calculated pair correlation function (black
line) and coordination number (red line) at temperatures of 3000 K
(solid line), 5000 K (dashed line), and 10000 K (dotted line). Inset
is the contour plot of the ionization degree as a function of
density and temperature.} \label{rdf}
\end{figure}

The plane-wave cutoff energy is selected to be 600.0 eV, so that the
pressure is converged within 5\% accuracy. $\Gamma$ point is
employed to sample the Brillouin zone in molecular dynamic
simulations, because EOS (conductivity) can only be modified within
5\% (15\%) for the selection of higher number of \textbf{k} points.
A total number of 128 atoms (64 deuterium molecules) is included in
a cubic cell, and over 300 (densities and temperatures) points are
calculated. The densities adopted in our simulations range from
0.167 to 0.9 g/cm$^{3}$ and temperatures between 20 and 50000 K,
which highlight the regime of principal Hugoniot. All the dynamic
simulations are lasted for 4 $\sim$ 6 ps, and the time steps for the
integrations of atomic motion are 0.5 $\sim$ 2 fs according to
different densities (temperatures). Then, the subsequent 1 ps
simulations are used to calculate EOS as running averages.
\begin{table}[!htbp]
\centering \caption{Coefficients $a_{ik}$ in expansion for the
internal energy $E$.}
\begin{tabular}{cccccc}
\hline\hline
$i$ & $a_{i0}$ & $a_{i1}$  & $a_{i2}$ & $a_{i3}$ & $a_{i4}$ \\
\hline
0 &     4.1065  &  -0.1111  &  13.4393  &  -6.7345  &   0.2532    \\
1 &     2.7497  &  -0.5432  &   2.6361  &  -9.6771  &   1.5115    \\
2 &    12.8700  &   1.6686  &  15.9184  &   8.8453  &  -3.3917    \\
3 &   -37.4966  & -18.1714  &  30.6954  &   2.8396  &   3.1274    \\
4 &     2.4465  &   3.1352  &   3.6421  &   5.3229  &  -1.2843    \\
\hline\hline
\end{tabular}
\label{coefficient_E}
\end{table}

In QMD simulations, zero point vibration energy
($\frac{1}{2}h\nu_{vib}$) and ionization energy (13.6 eV/atom) are
excluded, thus, the internal energy and pressure should be
corrected as follows:
\begin{equation} \label{tot_E}
E=E_{QMD}+\frac{1}{2}N(1-\alpha)E_{vib}+N\beta E_{ion},
\end{equation}
\begin{equation} \label{tot_pressure}
P=P_{QMD}+(1+\beta)\frac{\rho k_{B}T}{m_{D}},
\end{equation}
where $N$ is the total number of atoms for the present system, and
$m_{D}$ presents the mass of deuterium atom. The density and
temperature are denoted by $\rho$ and $T$ respectively, and $k_{B}$
stands for Boltzmann constant. $\alpha$ and $\beta$ are the
dissociation degree and ionization degree, while $E_{vib}$ and
$E_{ion}$ correspond to the zero point vibration energy and
ionization energy, respectively. $E_{QMD}$ and $P_{QMD}$ are
calculated from VASP. Various corrections to QMD simulations have
already been applied to model warm dense matter
\cite{PBX:Holst:2008,PBX:Lenosky:2000}, but contributions from
atomic ionization, which are particularly important at high
pressure, are still in absence. Ionization degree of aluminium under
extreme conditions has been successfully quantified through Drude
model \cite{PBX:Mazevet:2005}, however, the simple metallic model is
not suitable for the present system. Here, a new and effective
method in accounting for contributions to the EOS from molecular
dissociation and atomic ionization has been demonstrated.
\begin{table}[!htbp]
\centering \caption{Coefficients $b_{ik}$ in expansion for the
total pressure $P$.}
\begin{tabular}{cccccc}
\hline\hline
$j$ & $b_{j0}$ & $b_{j1}$  & $b_{j2}$ & $b_{j3}$ & $b_{j4}$ \\
\hline
0 &   41.9168  &  -1.2676    &  -8.5830  &  -55.5348   &  5.4594  \\
1 &  101.2582  &   6.0739    &  15.4631  &  -29.1378   & -8.2198  \\
2 &   60.8838  &  -0.6898    &   4.6233  & -124.8318   & 24.2147  \\
3 &  277.4649  & -10.0840    &  46.9646  & -233.0212   & -5.4740  \\
4 &    8.0324  &  -3.8221    &  -1.0263  &   30.4031   & -0.3011  \\
\hline\hline
\end{tabular}
\label{coefficient_P}
\end{table}

The dissociation degree is important in determining the internal
energy, from which EOS can be derived, especially at low
temperatures and the initial state on the Hugoniot curve. The
dissociation degree could be evaluated through the coordination
number:
\begin{equation} \label{fraction}
K(r)=\frac{N-1}{\Omega}\int_{0}^{r}4\pi r'^{2}g(r')dr',
\end{equation}
where $\Omega$ is the volume of the supercell. The coordination
number is a weighted integral over the pair correlation function
(PCF) $g(r)$ of the ions. The doubled value of $K$ at the maximum
of $g(r)$ ($r$ = 0.75 \AA), is equal to the fraction of atoms
forming molecules in the supercell. The sampled PCF and $K(r)$ are
labelled in Fig. \ref{rdf}. Fast dissociation of molecules emerges
at the temperature between 4000 and 7000 K, and a region featured
with $(\partial P/\partial T)_{V}<0$, which is not presented here,
is observed. Our results show that molecular deuterium can be
neglected above 15000 K due to thermal dissociation.

The ionization degree of the system can be evaluated through Saha
equation:
\begin{equation} \label{Saha}
\frac{\beta^{2}}{1-\beta}=\frac{2\Omega}{\lambda^{3}}\exp(-\frac{E_{ion}}{k_{B}T}),
\end{equation}
\begin{equation} \label{lamda}
\lambda=\sqrt{\frac{h^{2}}{2\pi m_{e}k_{B}T}},
\end{equation}
where only one level of ionization process is considered. In the
present formula, $m_{e}$ stands for the electron mass. As shown in
the inset in Fig. \ref{rdf}, the ionization of deuterium could be
neglected below 10000 K. At the temperature between 10000 and
50000 K, where the modeling of the principal Hugoniot of deuterium
is rather difficult, partially ionized warm dense fluid is formed,
and the ionization of atoms is of predominance in determining the
EOS.
\begin{figure}[!htbp]
\centering
\includegraphics[width=6.0cm]{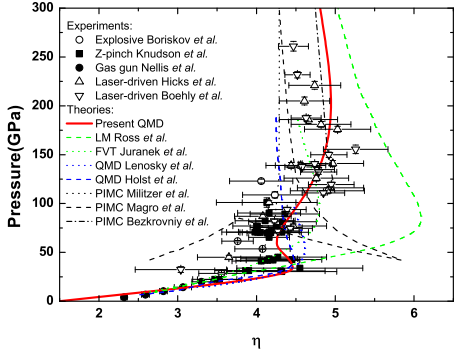}
\caption{(Color online) Simulated principal Hugoniot curve (solid
red curve). Previous data are also shown for comparison. Theories:
QMD results by Lenosky \emph{et al}. \cite{PBX:Lenosky:2000}
(dotted blue curve) and Holst \emph{et al}. \cite{PBX:Holst:2008}
(dashed blue curve); restricted PIMC simulations by Magro \emph{et
al.} \cite{PBX:Magro:1996} (dashed black curve), Militzer \emph{et
al.} \cite{PBX:Militzer:2000} (dotted black curve), and direct
PIMC by Bezkrovniy \emph{et al.} \cite{PBX:Bezkrovniy:2004}
(dashed dot black curve); the linear mixing model of Ross
\cite{PBX:Ross:1998} (dashed green curve), and the chemical model
FVT \cite{PBX:Juranek:2000} (dotted green curve). Experiments: gas
gun by Nellis \emph{et al.} \cite{PBX:Nellis:1983} (solid circle),
Z-pinch by Knudson \emph{et al.} \cite{PBX:Knudson:2004} (solid
square), explosives of Boriskov \emph{et al.}
\cite{PBX:Boriskov:2005} (open circle), laser-driven by Hicks
\emph{et al.} \cite{PBX:Hicks:2009} (up open triangle) and Boehly
\emph{et al.} \cite{PBX:Boehly:2004} (down open triangle).}
\label{hugoniot}
\end{figure}

Following Lenosky \emph{et al.} \cite{PBX:Lenosky:1997}, Beule
\emph{et al}. \cite{PBX:Beule:1999}, and Holst \emph{et al}.
\cite{PBX:Holst:2008}, we fit the internal energy and pressure by
expansions in terms of density (g/cm$^{3}$) and temperature
($10^{3}$ K). The corrected QMD data for internal energy (eV/atom)
can be expanded as follows:
\begin{equation} \label{E_expansion}
E=\sum_{i=0}^{4}A_{i}(T)\rho^{i},
\end{equation}
\begin{equation} \label{E_index}
A_{i}(T)=a_{i0}\exp[-(\frac{T-a_{i1}}{a_{i2}})^{2}]+a_{i3}+a_{i4}T.
\end{equation}
The total pressure given in GPa can be similarly expanded as $E$
with the expansion coefficients $b_{jk}$. The expansion coefficients
$a_{ik}$ and $b_{jk}$ for $E$ and $P$ (accuracy better than 5\%) are
summarized in Tab. \ref{coefficient_E} and Tab. \ref{coefficient_P},
respectively.

Based on the EOS, the principal Hugoniot curve can be derived from
the following equation:
\begin{equation} \label{EQ_hugoniot}
    (E_{0}-E_{1})=\frac{1}{2}(\frac{1}{\rho_{0}}-\frac{1}{\rho_{1}})(P_{0}+P_{1}),
\end{equation}
where the subscripts 0 and 1 refer to the initial and shocked
states. In our present simulations, the initial density $\rho_{0}$
is 0.167 g/cm$^{3}$ with the respective internal energy $E_{0}$ =
$-$3.28 eV/atom at $T_{0} = 20$ K. The pressure $P_{0}$ of the
starting point on the Hugoniot can be neglected compared to high
pressures of shocked states.

\begin{figure}[!htbp]
\centering
\includegraphics[width=6.0cm]{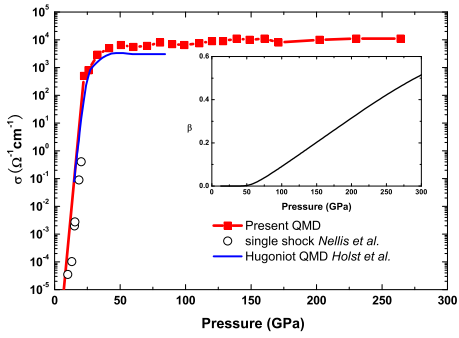}
\caption{(Color online) Calculated dc conductivity along the
Hugoniot curve (solid squares). Previous data
\cite{PBX:Nellis:1983,PBX:Holst:2008} are also shown for
comparison. Inset is the ionization degree along the Hugoniot.}
\label{conductivity}
\end{figure}

The principal Hugoniot is shown in Fig. \ref{hugoniot}, where
previous theoretical and experimental results are also provided for
comparison. At pressures below 100 GPa, our results indicate that
the principal Hugoniot experiences a local maximum compression ratio
of 4.5 around 40 GPa, which can be attributed to the dissociation of
molecules. The present Hugoniot agrees well with previous
experiments, such as gas gun \cite{PBX:Nellis:1983}, magnetically
launched flyer plates \cite{PBX:Knudson:2004}, and converging
explosives \cite{PBX:Boriskov:2005}. At the pressure between 40 and
100 GPa, the Hugoniot curve shows a stiff behavior, and the
compression ratio lies between 4.25 and 4.5. Meanwhile, the
ionization of atoms increases remarkably at $P$ $>$ 50 GPa (see the
inset in Fig. \ref{conductivity}) and consequently intenerates the
fluid. Thus, the combined effect of the molecular dissociation and
atomic ionization results in a local minimum of $\eta$ (4.25) along
the Hugoniot.

Recent high power laser-driven experiments
\cite{PBX:Boehly:2004,PBX:Hicks:2009} suggest that deuterium is
stiff ($\eta_{max} \approx 4.2$) below 100 GPa and become softer
($\eta \approx 4.5 \sim 5.5$) above 110 GPa, which can be described
by the present simulations. Our results indicate that molecular
deuterium can be neglected at this stage, and the atomic ionization
dominates the characteristic of the Hugoniot with $\eta$ lies
between 4.5 and 4.95 (maximum is reached at 200 GPa), which is
accordant with recent experiment \cite{PBX:Knudson:2009}. The
wide-range behavior of the Hugoniot is characterized by two stage
transitions---dissociation under low pressure and ionization at
higher pressure, and the present results show excellent agreement
with experimental ones. The Hugoniots from mere QMD simulations
hardly exceed 100 GPa \cite{PBX:Lenosky:2000} except for that of
Holst \emph{et al}. \cite{PBX:Holst:2008}, but $\eta$ does not
exceed 4.5 (at $P >$ 100 GPa). Although some PIMC simulations
\cite{PBX:Magro:1996,PBX:Bezkrovniy:2004} show five to six-fold
compressions, the simulated data are not yet comparable with
experiments. Due to the intrinsic approximations, no consistency has
been detected between our results and those of linear mixing model
\cite{PBX:Ross:1998} and chemical model FVT \cite{PBX:Juranek:2000}.

\begin{figure}[!htbp]
\centering
\includegraphics[width=6.0cm]{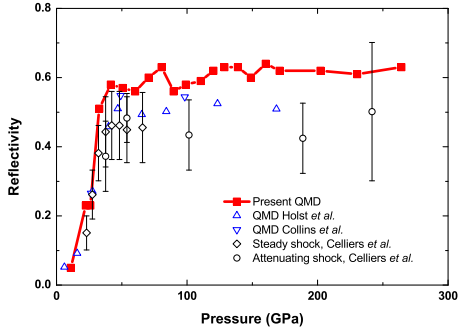}
\caption{(Color online) Calculated optical reflectivity of
wavelength 808 nm along the Hugoniot. Previous data
\cite{PBX:Celliers:2000,PBX:Holst:2008,PBX:Collins:2001} are also
plotted for comparison.} \label{fig_reflectivity}
\end{figure}

Let us turn now to see the nonmetal-metal transition by studying the
optical and conductive behaviors of the warm dense deuterium. The
real part of dynamic conductivity $\sigma_{1}(\omega)$ can be
evaluated through the following Kubo-Greenwood formula:
\begin{eqnarray} \label{real-conductivity}
\sigma_{1}(\omega)=&\frac{2\pi}{3\omega\Omega}\sum\limits_{\textbf{k}}w(\textbf{k})\sum\limits_{j=1}^{N}\sum\limits_{i=1}^{N}\sum\limits_{\alpha=1}^{3}[f(\epsilon_{i},\textbf{k})-f(\epsilon_{j},\textbf{k})]
\cr
&\times|\langle\Psi_{j,\textbf{k}}|\nabla_{\alpha}|\Psi_{i,\textbf{k}}\rangle|^{2}\delta(\epsilon_{j,\textbf{k}}-\epsilon_{i,\textbf{k}}-\hbar\omega),
\end{eqnarray}
where $i$ and $j$ summations range over $N$ discrete bands
included in the calculation. The $\alpha$ sum is over the three
spatial directions. $f(\epsilon_{i},\textbf{k})$ describes the
occupation of the $i$th band, with the corresponding energy
$\epsilon_{i,\textbf{k}}$ and the wavefunction
$\Psi_{i,\textbf{k}}$ at $\textbf{k}$. $w(\textbf{k})$ is the
$\textbf{k}$-point weighting factor.

From the calculated dynamic conductivity along the Hugoniot curve,
the dc conductivity $\sigma_{dc}$, which follows the static limit
$\omega$\texttt{$\rightarrow$}0 of $\sigma_{1}(\omega)$, is then
extracted and plotted in Fig. \ref{conductivity} as a function of
the Hugoniot pressure. As shown in Fig. \ref{conductivity},
$\sigma_{dc}$ increases rapidly with pressure up to 40 GPa towards
the formation of metallic state of deuterium, which agrees well with
the experimental measurements \cite{PBX:Nellis:1983}. Similar
tendency has also been found in the QMD simulations of warm dense
hydrogen \cite{PBX:Holst:2008}. When further increasing the Hugoniot
pressure, one finds from Fig. \ref{conductivity} that $\sigma_{dc}$
keeps almost invariant and the warm dense deuterium maintains its
metallic behavior. Here we address that the nonmetal-metal
transition is induced by gradual dissociation of molecules and
thermal activation of electronic states, instead of atomic
ionization, which is not observed until 50 GPa according to the
charge density distribution in the QMD simulations. Quantitative
analysis can be clarified through plotting the ionization along the
Hugoniot as shown in the inset in Fig. \ref{conductivity}.
Meanwhile, optical reflectivity, with the respective wavelength of
808 nm, is shown along the principal Hugoniot in Fig.
\ref{fig_reflectivity}, where good agreement has been achieved
between our present work and previous experiments
\cite{PBX:Celliers:2000}. The increase of reflectance (from 0.05 to
0.6) is observed, and this can be interpreted as a gradual
transition from a molecular insulating fluid to a partially
dissociated and metallic fluid at above 40 GPa.

In summary, we have performed QMD simulations to study the
thermophysical properties of deuterium under extreme conditions. The
Hugoniot EOS has been evaluated through QMD calculations and
corrected by taking into account the molecular dissociation
described by the coordination number $K(r)$ and the atomic
ionization described by Saha equation. The corrected Hugoniot has
shown good agreement with the experimental data in a wide range of
shock conditions, which thus indicates the importance of physical
picture of a two-stage transition, i.e., dissociation and
ionization. The principal Hugoniot reveals a local maximum
compression ratio of 4.5 at 40 GPa. With the increase of pressure,
$\eta$ reaches a maximum of 4.95 at about 200 GPa, and the
contribution from atomic ionization demonstrates softened character
of the Hugoniot. Smooth transition from a molecular insulating fluid
to an partially dissociated and metallic fluid are observed at 40
GPa. Our calculated optical constants along the Hugoniot have shown
excellent agreement with experiments. In addition, smooth fit
functions constructed in the present paper for the internal energy
and total pressure are expected to be useful for the future studies
of warm dense deuterium.

\begin{acknowledgments}
 This work was supported by NIFC.
\end{acknowledgments}


\end{document}